\documentclass[final,english]{bullsrsl}[2022/06/15]

\usepackage[latin1]{inputenc}
\usepackage[T1]{fontenc}
\usepackage{natbib} 
\usepackage{graphicx}
\begin{document}
\title{Photometric calibration of the 4m ILMT data}


\author[affil={1,2}, corresponding]{Bhavya}{Ailawadhi}
\author[affil={3,4}]{Talat}{Akhunov}
\author[affil={2,5}]{Monalisa}{Dubey}
\author[affil={2,5}]{Naveen}{Dukiya}
\author[affil={6}]{Sara}{Filali}
\author[affil={7, 8}]{Paul}{Hickson}
\author[affil={2}]{Priyanshi}{Kumari}
\author[affil={2}]{Gokul Singh}{Mehra}
\author[affil={2}]{Kuntal}{Misra}
\author[affil={9}]{Vibhore}{Negi}
\author[affil={2,10}]{Kumar}{Pranshu}
\author[affil={6,2}]{Jean}{Surdej}
\author[affil={2,10}]{Sarvesh Kumar}{Yadav}
\affiliation[1]{Physical Research Laboratory, Navrangpura, Ahmedabad, Gujarat-380009, India}
\affiliation[2]{Aryabhatta Research Institute of observational sciencES (ARIES), Manora Peak, Nainital, 263001, India}
\affiliation[3]{National University of Uzbekistan, Department of Astronomy and Astrophysics, 100174 Tashkent, Uzbekistan}
\affiliation[4]{ Ulugh Beg Astronomical Institute of the Uzbek Academy of Sciences, Astronomicheskaya 33, 100052 Tashkent, Uzbekistan}
\affiliation[5]{Mahatma Jyotiba Phule Rohilkhand University, Pilibhit Bypass Road, Bareilly 243006, Uttar Pradesh, India}
\affiliation[6]{Institute of Astrophysics and Geophysics, Li\`{e}ge University, All\'{e}e du 6 Ao$\hat{\rm u}$t 19c, 4000 Li\`{e}ge, Belgium}
\affiliation[7]{Department of Physics and Astronomy, University of British Columbia, 6224 Agricultural Road, Vancouver, BC V6T 1Z1, Canada}
\affiliation[8]{Outer Space Institute, The University of British Columbia, 325-6224 Agricultural Road, Van-
couver, BC V6T 1Z1, Canada}
\affiliation[9]{Kavli Institute for Astronomy and Astrophysics, Peking University, Beijing 100871, China}
\affiliation[10]{Department of Applied Optics and Photonics, University of Calcutta, Kolkata 700106, India}

\correspondance{bhavya@prl.res.in}
\date{13th October 2020}
\maketitle

\begin{abstract}
The International Liquid Mirror Telescope (ILMT) is a 4-meter survey telescope continuously
observing towards the zenith in the SDSS g', r', and i' bands. This survey telescope is designed
to detect various astrophysical transients (for example, supernovae) and very faint objects like
multiply-imaged quasars and low surface brightness galaxies. Ensuring accurate and stable calibration of supernova (SN) and quasar measurements is essential for studying their intrinsic variability and long-term brightness evolution. In this work, we present a dedicated calibration approach tailored for ILMT point-like object observations obtained in Time Delayed Integration (TDI) mode. Using ZTF reference sources cross-matched with the Pan-STARRS1 catalog, we derive robust photometric zero-points along the ILMT scan. This calibration procedure enables the production of consistent, high-precision SN and quasar light curves, suitable for variability studies, structure-function analysis, and future reverberation-mapping applications.
\end{abstract}


\section{Introduction}
The 4-m International Liquid Mirror Telescope (ILMT; \citealt{Surdej2018, Surdej2025}), located at Devasthal (ARIES, India), is the first liquid mirror facility dedicated exclusively to time-domain astronomical surveys in South Asia. Operating in the SDSS g', r', and i' bands, the telescope continuously scans the zenith and records a $\sim$22-arcmin-wide declination strip of the sky. Because the same region is observed every night, the ILMT naturally produces repeated measurements of thousands of sources, making it highly effective for detecting and characterizing transient and variable phenomena such as supernovae, variable stars, and quasars \citep{Pranshu2025}. These observations are acquired using a charge-coupled device (CCD) in Time Delay Integration (TDI) mode, which tracks the apparent motion of celestial sources caused by Earth's rotation. However, the TDI acquisition method, combined with changing atmospheric conditions, introduces systematic variations in the recorded fluxes, meaning that the raw instrumental magnitudes are not directly comparable to any standard photometric system. To derive meaningful scientific information such as brightness evolution, color indices, and physical properties of the detected sources, it is therefore essential to transform these instrumental measurements onto a stable, well-calibrated photometric scale.

Photometric calibration ensures that the measured brightness of celestial objects reflects their true, standardized flux, enabling meaningful comparisons across different nights, filters, and observatories. In this work, we employ Zwicky Transient Facility (ZTF) reference stars, whose calibrated magnitudes are derived from the Pan-STARRS1 (PS1; \cite{Chambers}) catalog, to establish photometric zero-points for ILMT observations. ZTF standards are chosen because they are well-tested and widely used in generating accurate light curves for many transient and variable star studies. Their proven reliability helps ensure that ILMT magnitudes are placed on a consistent and stable photometric scale. This pipeline's calibrated light curves allow for reliable source classification, effectively distinguishing between rapidly changing transients and periodic variable stars. In this article, the methodology and steps adopted for calibration purpose are outlined in Section \ref{method}. Some initial results of the calibration are described in Section \ref{results}. The work is summarised in Section \ref{summary}.

\section{Methodology}
\label{method}
This section outlines the steps and techniques used to extract the calibrated magnitudes of the point-like sources from the instrumental magnitudes evaluated through photometry \citep{ailawadhi}. The sequence of steps followed is described below.

\begin{itemize}
    \item Selecting stars located within the ILMT's 22-arcmin declination strip, ensuring that only sources actually observed during the zenith scan are used for calibration.
    
    \item Cross-matching ILMT detections with ZTF reference stars within a 1-arcsec radius to obtain a clean set of common calibrators with reliable standard magnitudes.
    
    \item Computing zero-point corrections for each ILMT frame using the matched calibrators from the same field, thereby accounting for atmospheric and instrumental variations.

    \item Applying this correction to the instrumental magnitudes to obtain the calibrated magnitudes.
\end{itemize}

\section{Results}
\label{results}
\begin{figure}[ht] \centering \includegraphics[width=1\textwidth]{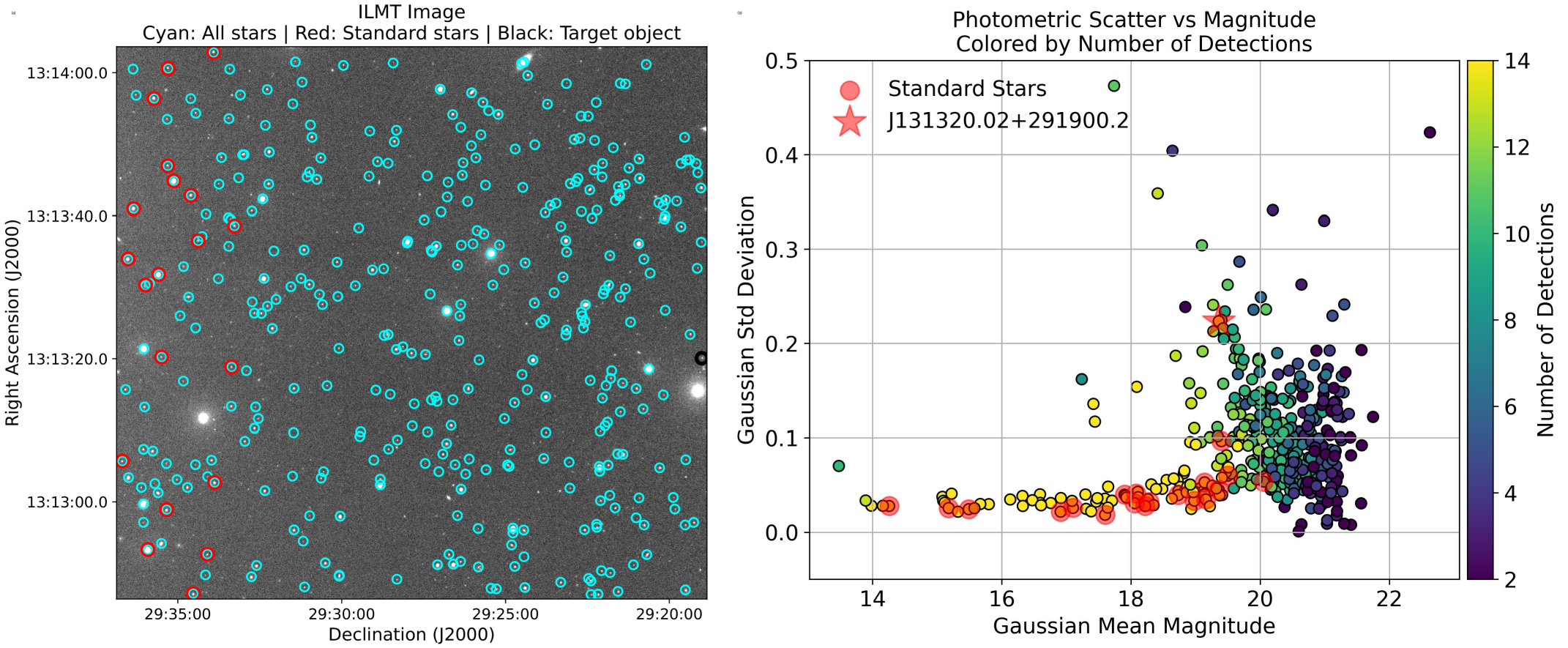} 
\begin{minipage}{12cm} \small \caption{Left panel of the Figure shows an ILMT field segment displaying quasar J131320.02+291900.2 (black) at the extreme right, along with several detected sources (cyan) and ZTF calibrator stars (red) used for photometric calibration. Right panel shows the variation of photometric standard deviation with respect to mean magnitude for all detected sources which are color coded by the number of detections. Brighter stars exhibit lower scatter, confirming the stability of the selected ZTF calibrators (red circles) used for magnitude calibration, while the red star marks the quasar.} \label{fig:quasar_1} \end{minipage}
\end{figure}



The frames containing the quasar J131320.02+291900.2 were calibrated using the methods described in Section~\ref{method}, selecting those with a starting LST of 13h,10m to ensure its inclusion within the $\sim$17-arcmin field. The left panel of Figure~\ref{fig:quasar_1} presents the corresponding ILMT frame segment observed on 11 Feb 2024, showing the quasar (in black), the ZTF calibrator stars used for photometric calibration (in red), and other detected sources exceeding the $3\sigma$ threshold.

Using a similar technique, we analyzed all ILMT frames with a starting LST of 13h\,10m obtained between April 2023 and April 2025. Over this two-year period, a total of 8, 16, and 25 frames were available in the \textit{g}, \textit{r}, and \textit{i} spectral bands, respectively. For the \textit{r}-band analysis, each detected source was tracked across all available nights, and the standard deviation of its calibrated magnitudes was plotted against its mean magnitude, as shown in the right panel of Figure~\ref{fig:quasar_1}. The data points are color-coded according to the number of nights in which each object was detected, with yellow indicating sources detected on most of the nights and blue indicating those detected on the fewest. The distribution shows that stars (mean magnitude $<19$) exhibit a scatter of approximately 0.05 mag, several of which correspond to ZTF calibrators (marked with red circles). On the fainter side, the scatter increases. For the quasar, a Gaussian fit to the measured magnitudes gives a mean magnitude of 19.33 mag and a standard deviation of 0.22 mag in the \textit{r} band (as shown with the red star).  

\begin{figure}
    \centering
    \includegraphics[width=0.85\textwidth]{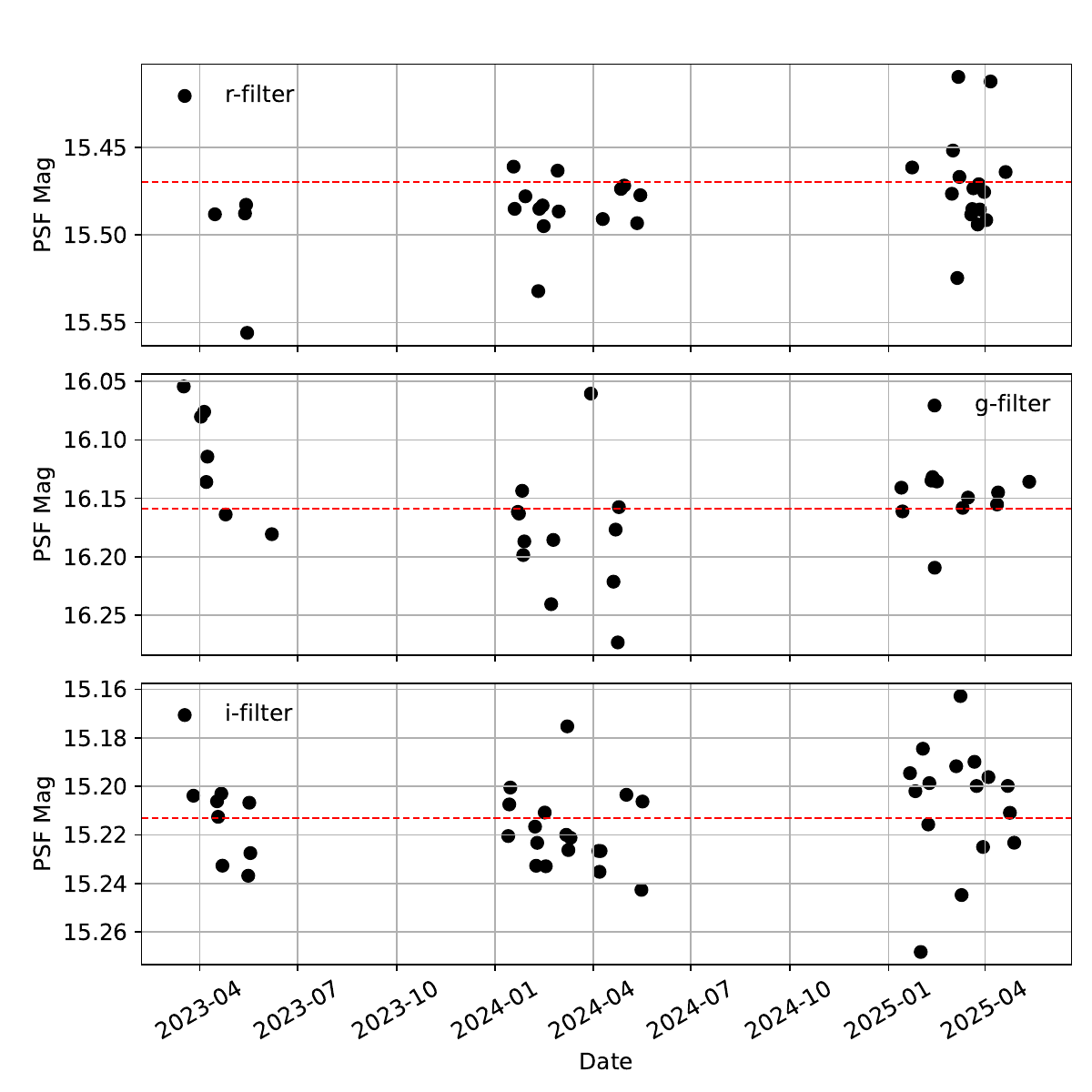}
    \begin{minipage}{12cm} \small
    \caption{Calibrated light curves of the ZTF standard star J131331.78+293534.62 in the \textit{g}, \textit{r}, and \textit{i} bands. The dashed line marks the adopted reference magnitude.}
    \label{fig:standard_star}
    \end{minipage}
\end{figure}

\begin{figure}[ht]
    \centering
    \includegraphics[width=0.85\textwidth]{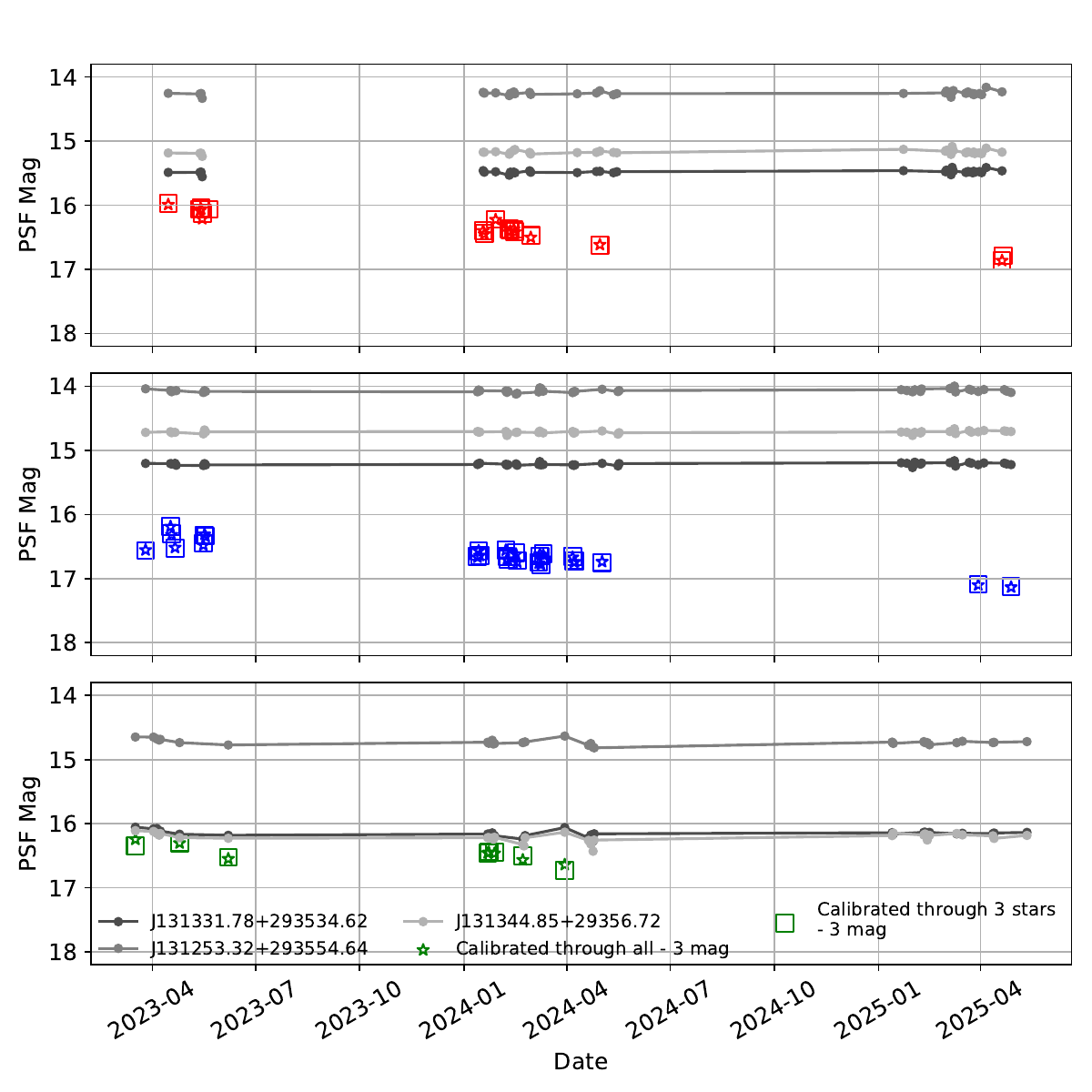}
    \begin{minipage}{12cm} \small
    \caption{Calibrated light curve of the quasar J131320.02+291900.2. Square points denote calibration using the three most stable standard stars, while star-shaped points represent calibration using all field stars, demonstrating the robustness and consistency of the photometric calibration.}
    \label{fig:quasar}
    \end{minipage}
\end{figure}

To further assess the robustness of the photometric calibration across all observing nights, we examined the long-term light curves of both the selected ZTF standard stars and the quasar. Figure~\ref{fig:standard_star} presents the calibrated light curves of the ZTF standard star J131331.78+ 293534.62 in the \textit{g}, \textit{r}, and \textit{i} bands. The dashed horizontal line marks the adopted reference magnitude, demonstrating that the nightly calibrated magnitudes closely cluster around this value, reflecting the stability and reliability of the chosen calibrators. Figure~\ref{fig:quasar} shows the light curve of the quasar J131320.02+291900.2. Here, the square points represent magnitudes calibrated using the three most stable standard stars, while the star-shaped points correspond to calibrations obtained using all available field stars. The excellent agreement between these two approaches highlights the consistency of our method and confirms that the adopted calibration strategy using all calibrators in the field yields robust quasar magnitudes across the full two-year observing period.

\section{Summary and Conclusions}
\label{summary}

In this work, we developed an automated Python-based pipeline to perform reliable photometric calibration of International Liquid Mirror Telescope data using stable reference stars from the Zwicky Transient Facility. The long-term consistency of these ZTF calibrators ensures accurate transformation of instrumental magnitudes onto a standard scale. The calibrated light curves produced by this method enable robust identification and characterization of astrophysical transients and variable stars. Overall, the workflow provides a scalable and dependable approach for current and future time-domain surveys requiring uniform and precise photometric measurements.

\begin{acknowledgments}
This work is a result of the Belgo-Indian Network for Astronomy and Astrophysics (BINA),
which promotes astronomical collaborations between Indian and Belgian partners and received
financial support from the International Division of the Department of Science and Technology (DST, Government of India) and the Belgian Federal Science Policy Office (BELSPO,
Government of Belgium) during the period 2016-2023 through different bilateral projects.
The 4-m International Liquid Mirror Telescope (ILMT) project results from a collaboration between the Institute of Astrophysics and Geophysics (University of Li\`{e}ge, Belgium), the Universities of British Columbia, Laval, Montreal, Toronto, Victoria, and York University, and Aryabhatta Research Institute of Observational Sciences (ARIES, India). The team acknowledges the contributions of ARIES's technical and scientific staff. JS wishes to thank Service Public Wallonie, F.R.S.-FNRS (Belgium), and the University of Li\`{e}ge, Belgium, for funding the construction of the ILMT. PH acknowledges financial support from the Natural Sciences and Engineering Research Council of Canada, RGPIN-2019-04369. PH, SF, and JS thank ARIES for hospitality during their visits to Devasthal. We sincerely acknowledge Anna Pospiezalska for her significant contributions to ILMT functioning. MD acknowledges the Innovation in Science Pursuit for Inspired Research (INSPIRE) fellowship award (DST/INSPIRE Fellowship/2020/IF200251) for this work. KP acknowledges the support from the Erasmus+ Programme of the European Union for a research visit to the Institute of Astrophysics and Geophysics, University of Li\`ege, Belgium. TA thanks the Ministry of Higher Education, Science, and Innovations of Uzbekistan (grant FZ-20200929344). JS and KM acknowledge the assistance received from the ANRF SERB-726 VAJRA Faculty Scheme, India. KM acknowledges the support from the BRICS grant DST/ICD/BRICS/Call-5/CoNMuTraMO/2023 (G) funded by the DST, India.    
\end{acknowledgments}

\begin{furtherinformation}

\begin{orcids}
\orcid{0009-0000-1020-9711}{Bhavya}{Ailawadhi}
\orcid{0000-0003-1637-267X}{Kuntal}{Misra}
\orcid{0000-0002-7005-1976}{Jean}{Surdej}
\end{orcids}

\begin{authorcontributions}
This work results from a long-term collaboration to which all authors have made significant contributions.
\end{authorcontributions}

\begin{conflictsofinterest}
The authors declare no conflict of interest.
\end{conflictsofinterest}

\end{furtherinformation}

\bibliographystyle{bullsrsl-en}

\bibliography{extra}

\end{document}